# INVERTING THE SOUND SPEED PROFILE FROM MULTI-BEAM ECHO SOUNDER DATA AND HISTORICAL MEASUREMENTS - A SIMULATION STUDY

Y Gourret University of South-Eastern Norway, Horten, Norway
T Brander University of South-Eastern Norway, Horten, Norway
KT Hjelmervik University of South-Eastern Norway, Horten, Norway

## 1 INTRODUCTION

The ocean's lack of transparency offers plenty of opportunities for illegal trade enterprises[1], terrorism, and covert military operations. Unmanned surface vehicles and autonomous underwater vehicles are possible disruptive technologies that might counter these threats[2]. The lack of human decision makers puts strong requirements on the autonomy of the vessels. Also, the inherent compactness of these vessels imposes strong restrictions on the amount and size of installed hardware.

For manned sonar operations, sound speed probes are used to determine the local sound speed profile, which is then used to estimate the sonar performance. For unmanned vessels, the extra payload for sound speed probes should be avoided. Ocean models provide an alternative source of information on the underwater environment. However, their predictions are still unable to provide the level of detail and accuracy required for small-scale sonar operations in littoral environments. Climatological data[3] and other statistically derived oceanographic data[4,5] may also provide information on the present sound speed, but fail to account for local spatio-temporal effects.

Through-the-sensor techniques is an alternative means of assessing the environment. Sound speed inversion from multi-beam echo sounder (MBES) data is addressed in several studies where the main goal is to improve topographic measurements[6,7,8,9]. Common for all these studies is that both the sound speed and the topography are assumed unknown in the inversion, and that they focus on the accuracy of the topographic measurements, rather than the quality of the sound speed profile (SSP). Hjelmervik[10] proposed inverting the sound speed from single-beam echo sounder (SBES) data and direct measurements of the sound speed at limited depths in an area with known topography. A major limitation in this work is that an SBES yields information on the average SSP only, not its structure. Keyzer et al.[9] inverts the SSP from MBES data in a similar manner using an empirical orthogonal function (EOF) representation for the SSPs. The advantage of using MBES data is that beams directed off the vertical will be refracted and thus contain information on the sound speed gradients throughout the depths. However, the cost function used by Keyzer et al. does not include any constraints that ensure that the resulting profile is physical.

We invert the SSP from MBES data in an environment with known topography. The SSPs are represented using EOFs, such that the proposed SSPs can be modelled by a limited number of coefficients[10]. An acoustic raytracer models the travel time from the transmitter to the bottom and back to the receiver for proposed SSPs. The cost function compares the modelled and measured travel times for all beam angles, but also includes *a priori* knowledge on the SSPs through Tikhonov-type regularization[11]. We assess the validity and performance of our method through simulations, e.g. to explore its sensitivity to errors in the measured travel times. Our work thereby expands on former work[10,9], as we include both MBES data to capture the vertical structure in the sound speed profile as well as constraining the EOF space explored by the method to physically feasible regions through the use of regularization. We also show that the problem is non-unique[11] without regularization.

# 2 THEORY

Recovering the sound speed is a travel time tomography inverse problem[12], since the travel time depends on the SSP[13]. Each measurement yields a set of travel times corresponding to different beams processed from the received data on the MBES. We assume range-independent sound speed.

As a simple example, consider a piecewise constant SSP, where the $i$th vertical layer has a constant sound speed of $c_i$ and a depth of $\Delta Z_i$. Consider a receiving beam directed with an initial angle $\theta_1$ to the vertical at the receiver. Assume a flat bottom with a seafloor depth $z_b$. The arrival time of a signal arriving in this beam following a reflection off the sea floor, is then given by

$$t = \sum_{i=1}^{N} \frac{\Delta\sigma_i}{c_i} = \sum_{i=1}^{N} \frac{\Delta z_i}{c_i \cos\theta_i}. \tag{1}$$

where $\sigma_i$ is the length of the trajectory through the $i$th layer. The latter expression follows from the trajectory following a straight path due to the constant sound speed in each layer. By taking into account Snell-Descartes law we then have

$$t = \sum_{i=1}^{N} \frac{\Delta z_i}{c_i\sqrt{1-c_i^2 p^2}} \quad \text{where } p = \frac{\sin\theta_1}{c_1} = \frac{\sin\theta_i}{c_i}. \tag{2}$$

The expression does not depend on the order of the vertical layers. Thus two layers may be switched without impacting the travel times. Hence, the mapping from SSP to travel time is not injective, since there exist several different SSPs that give the same travel times. This ambiguity causes the non-uniqueness of the inverse problem.

Since we have plenty of physical information about the SSP, we can use *a priori* information to deal with the lack of uniqueness. Not all vertical sound speed structures are physical even if they provide solutions to the inverse problem, and even if they are physically possible, they will have different associated probabilities of occurring.

To structure the historical sound speed measurements, the SSPs are represented by a weighted sum of EOFs following the steps of Preisendorfer[14]. The EOFs are determined from a set of historical sound speed profiles from an area of limited geographical extent and a limited part of the year.

We use the vector $\mathbf{x} = [x_1, x_2, ..., x_K]^T$, where $x_k$ are the EOF coefficients, to write the SSP as

$$\boldsymbol{c}(\mathbf{x}) = \widehat{\boldsymbol{c}} + U\mathbf{x}. \tag{3}$$

The average SSP, $\widehat{c}$, may be estimated from historical SSPs, which are first interpolated to a common depth grid with $K$ depth steps. The matrix $U$ then becomes a $K \times K$ matrix containing the EOFs estimated from the historical SSPs.

The coefficient vector $\mathbf{x}_n$ of the $n$th historical sound speed profile $\boldsymbol{c_n}$ may be found by

$$\mathbf{x}_n = U^T(\boldsymbol{c_n} - \widehat{\boldsymbol{c}}). \quad (4)$$

Given a large amount of historical measurements and statistical independence between different coefficients, the probability density functions of each coefficient, $f_{xk}(x)$, may be derived. Assuming Gaussian behaviour, the distribution for the EOF coefficients $\mathbf{x}$ then becomes

$$f_{\mathbf{x}}(\mathbf{x}) = \prod_{k=1}^{K} \frac{1}{\sqrt{2\pi}\sigma_k} \exp\left(-\frac{x_k^2}{\sigma_k^2}\right), \quad (5)$$

where $\sigma_k$ is the standard deviation of the $k$th EOF coefficient and may be estimated from the historical dataset.

The number of EOFs corresponds to the number of depth steps $K$. Most of the sound speed variance is captured by the first few EOFs. The lower order of EOFs represent the bulk of the information[14]. We therefore choose $N_{EOF}$ as a cutoff for the number of EOFs used and thereby reduce the matrix $U$ to a size of $K \times N_{EOF}$ and the vector $\mathbf{x}$ will only have $N_{EOF}$ elements.

**Assumptions**

We consider an environment with a flat bottom of known depth. The position and altitude of the MBES is also known. The MBES records $N_{beam}$ infinitely narrow beams, and the beam directions are perfectly known. The error on travel time is independent and Gaussian distributed with standard deviation $\sigma_\tau$. This error accounts for the measurement error at the receiver due to the imperfect detection of a received signal embedded in noise, but also for imperfect knowledge of the seafloor depth.

In this simulation study we limit the error source to an error in the received travel time. In reality, the results will also be influenced by a plethora of other sources of error for different parameters, such as angular errors and uncertainty in the forward model, but this is out of scope for this study.

## 2.1 Cost Function

Let $t(\mathbf{x})$ be a function of the EOF coefficients $\mathbf{x} = [x_1, x_2, ..., x_K]^T$. This corresponds to the modelled travel times which equation (1) is a simplified model for. Here we employ an acoustic raytracer to estimate the travel time for increased accuracy.

The proposed cost function minimizes the difference between measured ($T_{measure}$) and the modelled ($t(\mathbf{x})$) travel times. The cost function also takes into account the historical information by including a regularization term containing the log-likelihood function for the EOF coefficients $\mathbf{x}$. We obtain

$$CF(\mathbf{x}) = \frac{\|T_{measure} - t(\mathbf{x})\|^2}{N_{beam}} + \alpha \sum_{i}^{N_{EOF}} \frac{x_i^2}{\sigma_i^2}. \quad (6)$$

To solve the minimization problem, we use the Gauss-Newton method. The problem is non-linear, but may empirically be solved for small variations. The mapping $\mathbf{x} \mapsto t(\mathbf{x})$ is continuous for a flat bottom and the angles and SSPs where the rays are not refracted back to the surface, but the differentiability assumed by Gauss-Newton is unknown. The validity of the proposed method is limited to flat or slowly varying seafloors. For complex seafloors small variations in the ray trajectory may result in large variations in travel time.

The value $\alpha > 0$ is a Tikhonov-type regularization parameter[11], that ensures a balance between historical sound speed measurements and the travel time measurements. A large $\alpha$ gives more weight to the historical information, while a low value gives more weight to the acoustic measurements. Generally, selection of the regularization parameter is a difficult task[11]. Here we use a simple neural network (NN) to determine a good choice for $\alpha$.

To work, a NN should be trained with many examples. The training dataset used to generate the EOFs may be too small for this training. However, by randomizing the coefficients $x_i$ according to the Gaussian distribution, we can recreate synthetic SSPs cf. eq. (3). This allows the generation of arbitrary amounts of synthetic examples to train our NN.

Figure 1 gives an overview of how the neural network works. We invert the SSP for $N_{inv}$ different values of $\alpha$ (on a logarithm scale). For each inversion, we get travel time errors and EOF coefficients $\mathbf{x}$, that are normalized respectively by the formula (a) and (b) shown in Fig. 1. We define $A_i$ as the normalized information coming from the inversion, cf. formula (c) in Fig. 1. The neural network is a multilayer perceptron trained to output the SSP error for $\alpha_i$ with list $A_{[i-k,i+k]}$ as input, with $k$ as the size of the operative window. This operation will be repeated for every value of $\alpha$. The NN requires a fixed number of inputs, so at boundaries we duplicate $A_i$, if $i \notin [1, N_{inv}]$ then $A_i = A_1$ or $A_{N_{inv}}$. Even if not indicated on the figure, the number of beams is also an input of the NN.

The final $\alpha$ used for the inversion is the $\alpha_i$ that gives the lowest response from the NN.

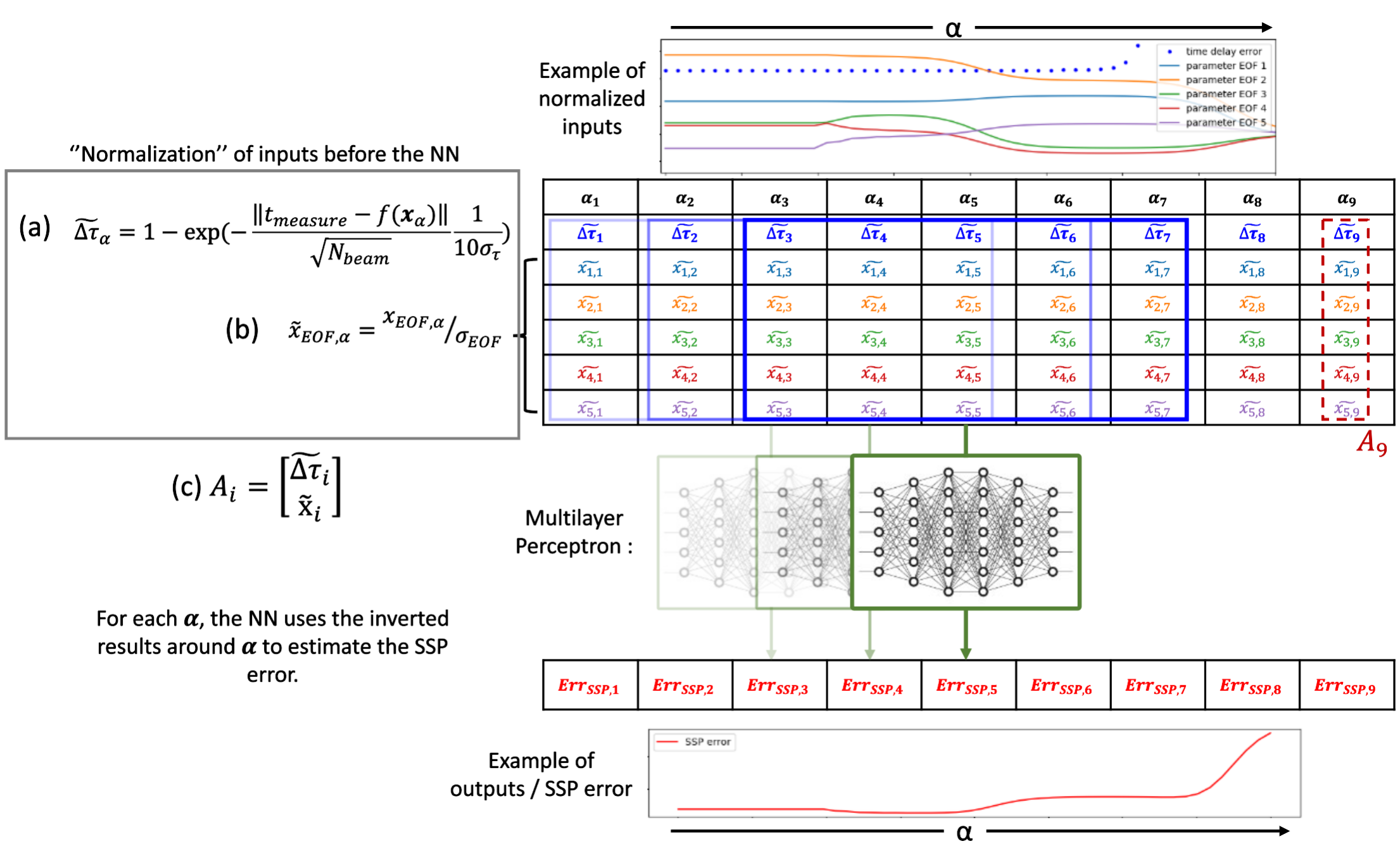

***Figure 1: An overview of the neural network. The blue border represents the operative window.***

# 3 DATASET

The MBES dataset is generated through ray-based simulations. The simulations employ a large set of sound speed profiles measured in the Norwegian trench on the West coast of Norway from 1990 to 2010. The sound speed measurements were taken from a validated dataset[15].

We divided the data into two sub-datasets; a training and a testing dataset. Both are geographically cropped according to a bounding box defined by $[59.849°, 62.092°]$ in latitude and $[2.924°, 4.990°]$ in longitude (South-West coast of Norway). The profiles are also cropped to a depth of 300 m. The training data (151 SSPs) were measured in April in the years from 1990 and 2000, while the testing data (111 SSPs) were measured in April in the years between 2001 and 2010. The objective is to invert the test SSPs based on historical information provided by the training data.

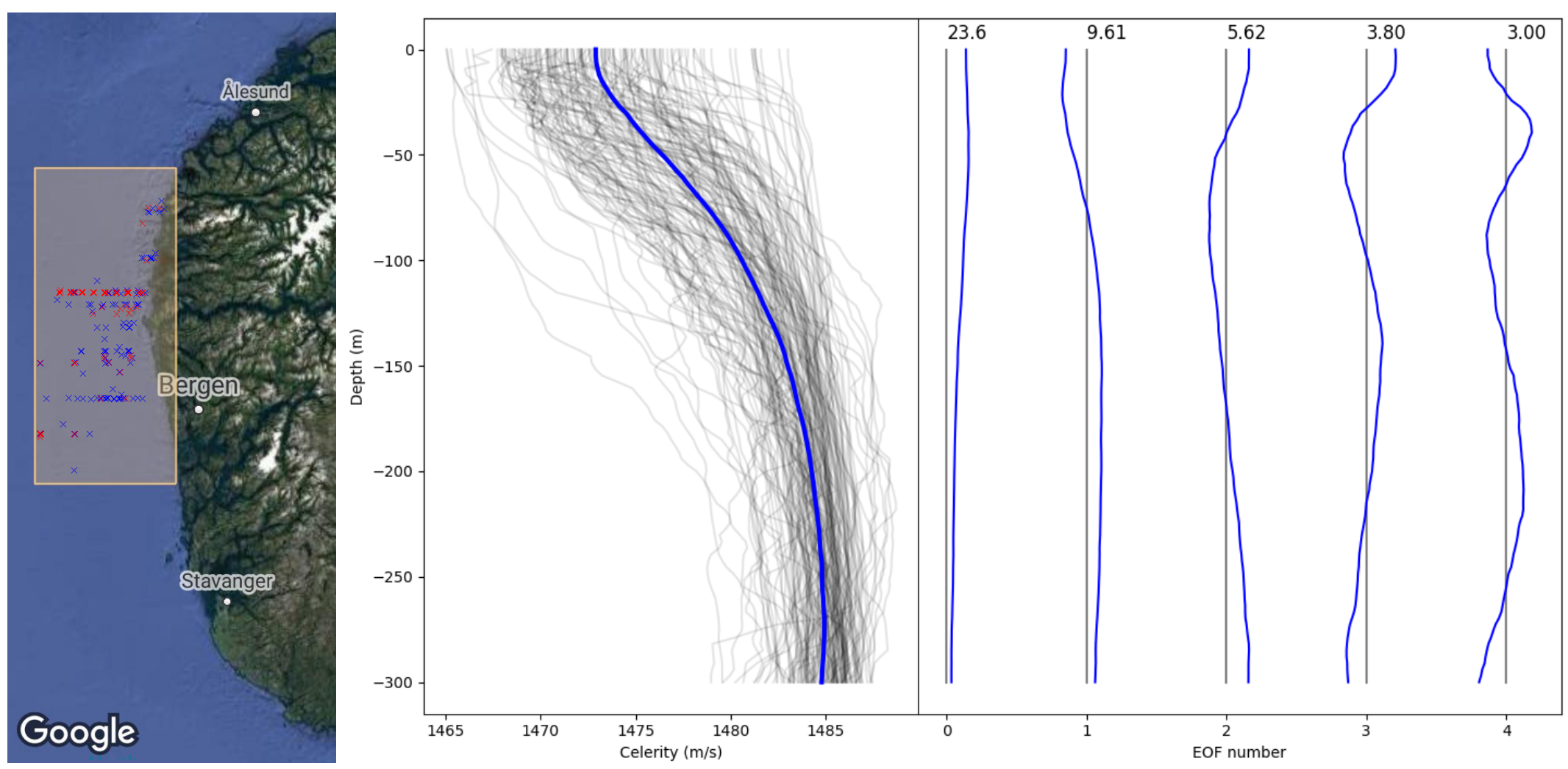

***Figure 2: Left: Dataset area in orange, training data in blue, testing data in red. Right: Training dataset information, average SSP and 5 first EOFs.***

# 4 RESULTS AND DISCUSSION

The proposed method was applied on a simulated raytrace based on the test dataset. Figure 3 presents the statistics for the errors of the inverted SSPs along with a single representative example of an inverted SSP. The main bulk of errors lies around 0.7 – 0.9 m/s on average over the entire depth. By comparison, the mean SSP error between the test dataset and the average SSP $\widehat{c}$ (see eq. 3) is 3.22 m/s.

Tab. 1 shows the results of a sensitivity analysis for different parameters. The central column (green) lists the parameters used to produce the results shown in Fig. 3. The table presents the average RMS error of the inverted SSPs of the testing dataset for different parameters. This permits us to have an overview of the impact of every parameter. For comparison the average error obtained from simply comparing the average SSP in the training data set to all the test SSPs is 3.22 m/s, and 2.55 m/s when using the average SSP from the test data set. Lacking the historical measurements, a natural choice of SSP to use during a sonar operation could be a climatological estimate from for instance the World Ocean Atlas[16]. This would result in an error of 2.62 m/s, which is still twice as high as the maximum error observed in the parameter variations shown in Tab. 1.

| **Parameters** | | | | **Variations** | | | |
|---|---|---|---|---|---|---|---|
| **Number of beams×pings** | | 100 | 300 | 500 | 700 | 900 | |
| RMS error (m/s) | | 0.92 | 0.87 | 0.83 | 0.80 | 0.79 | |
| **Swath angle width (deg)** | | 100 | 110 | 120 | 130 | 140 | |
| RMS error (m/s) | | 1.1 | 0.96 | 0.83 | 0.77 | 0.75 | |
| **Number of EOFs** | | 3 | 4 | 5 | 6 | 7 | 8 |
| RMS error (m/s) | | 0.94 | 0.85 | 0.83 | 0.81 | 0.83 | 0.82 |
| **Spatial error (cm)** | 10 | 4 | 2 | 1 | 0.5 | 0.25 | |
| RMS error (m/s) | 1.36 | 1.1 | 0.93 | 0.83 | 0.79 | 0.78 | |

***Table 1: RMS error for different variations of parameters used in the overall method. The central column, in green, corresponds to the results shown in fig. 3. The spatial error $\sigma_x$ is here given by $\sigma_x = c\sigma_t/2$.***

The first parameter considered is the amount of beams times the amount of pings. This is a combination of the amount of beams along the swath and the amount of pings accumulated before inverting the sound speed. This also includes extra measurements accumulated by exploiting that the bottom might be detected at different time samples within the same beam. However, it represents the number of *independent* measurements. An increase here results in a decreasing SSP RMS error. This is expected since an increased amount of travel time measurements should result in an improved estimate of the SSP, in much the same way as repeated, independent measurements of a random value will increase the accuracy of an estimate for that value.

The second parameter is the swath angle width of the MBES. For narrow swath angle, the sonar measurements are all directly below the ship. This only provides information on the average SSP, as a wave travelling directly downwards in a range-independent environment remains unrefracted and has a travel time dependent on the average sound speed and seafloor depth only. By increasing the opening angle, we increase the incidence angle of the sound wave to the water layers at the side of the MBES. This increases the refraction effects providing us more information for the inversion of the SSP. As we observe on Tab. 1, the accuracy of the inversion increases with the opening angle. The achieved gain from widening the opening angle is limited to the spectrum of angles that result in bottom interaction. Depending on the sound speed profile, angles close to the horizontal can be reflected by water layers, and thus there might be no bottom interaction.

The third parameter is the number of EOFs used for the inversion. As we can see, the SSP error decreases until six EOFs are used, but for seven EOFs, the SSP error increases. Thus, increasing the number of EOFs further does not increase the quality of the inversion. The number of EOFs should be chosen carefully. The quality of the EOFs depends on the number of SSPs used to obtain them. Higher-order EOFs require more SSPs to be determined accurately, and they are more likely to change over the years than lower-order EOFs.

The last parameter is the range accuracy of the MBES. Smaller spatial error leads to greater accuracy of sound speed recovery, as expected.

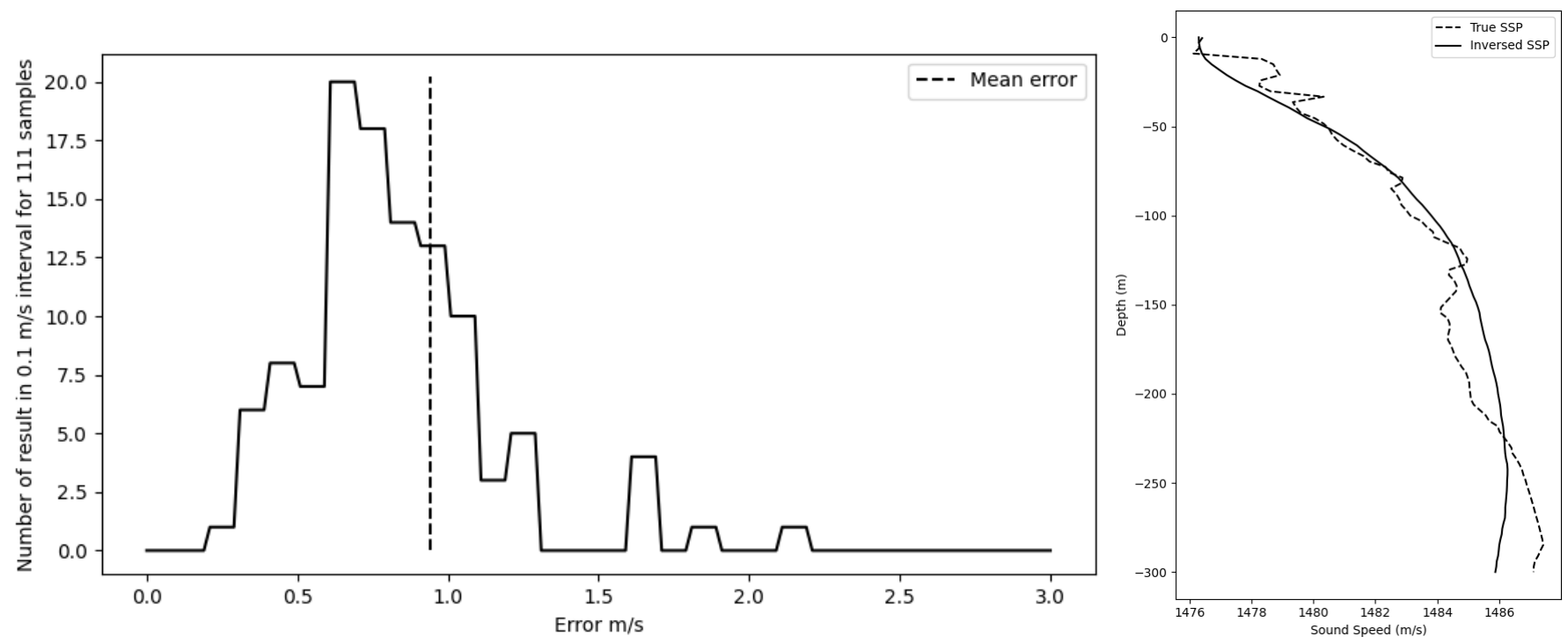

***Figure 3: Left: Histogram of the RMS error for the inverted SSPs for the test dataset. The average SSP RMS error of 0.83 m/s is shown as a vertical dashed line. Right: One example of a true and inverted SSP with an error of 0.83 m/s.***

# 5 SUMMARY

We showed with an elementary example that the inverse problem of recovering an SSP from travel time measurements is generally non-unique, leading to an ill-posed inverse problem. However, by using EOFs from historical data and a Tikhonov-type regularization we showed that we are able to recover SSPs with an accuracy of 0.83 m/s for an MBES system with a maximum error in its spatial estimates of 1 cm. Larger spatial errors translate into increased errors in the SSP up to 1.36 m/s for spatial errors of 10 cm, which still outperforms the climatological alternative from the World Ocean Atlas[16] by a factor close to 2.

# ACKNOWLEDGMENTS

This research has been supported by the European Commission within the context of the project SMAUG (Smart Maritime and Underwater Guardian), funded under EU Horizon Europe Grant Agreement 101121129.

We would like to thank Ailin Brakstad for preparing the sound speed data used in this study.